\documentstyle[12pt]{article}

\def\p{{\rm {\bf p}}}
\def\q{{\rm {\bf q}}}

\def\exp{{\rm exp}}
\def\g{\mbox{\boldmath$\gamma$}}

\begin{document}

\vspace{1.5cm}

\begin{center}

{\bf ONE-PARTICLE AND COLLECTIVE ELECTRON SPECTRA IN HOT AND DENSE QED
AND THEIR GAUGE DEPENDENCE}

\vspace{1.5cm}

{\bf O.K.Kalashnikov}
\footnote{ E-mail address:kalash@td.lpi.ac.ru}

Theoretical Physics Division,

P.N.Lebedev Physical Institute,

Russian Academy of Sciences,

117924 Moscow, Russia.

\vspace{2.5cm}

{\bf Abstract}

\end{center}

The one-particle electron spectrum is found for hot and dense QED and
its properties are investigated in comparison with the collective
spectrum. It is shown that the one-particle spectrum (in any case its
zero momentum limit) is gauge invariant, but the collective spectrum,
being qualitatively different, is always gauge dependent. The
exception is the case $m,\mu=0$ for which the collective spectrum long
wavelength limit demonstrates the gauge invariance as well.

\newpage

\section{Introduction}
At present a problem exists to establish whether the effective electron
mass in statistical QED (the same as in QFT) is gauge invariant or
this mass depends on the calculational schemes, in particular on the
gauge fixed or on the chosen regularization procedure to treat the
infrared and ultraviolet divergencies. This problem is rather
complicated and up today is not solved finally although many attempts
were made starting from the fiftieth years [1,2]. In these pioneer
papers the gauge dependence of the electron Green function was
investigated at first in QFT and then in statistical QED [3]. It
was shown that the longitudinal part of the gauge field generated
only the phase factor for the electron Green function and its gauge
dependence was separated but, unfortunately, only in the coordinate
space. In the momentum space where the physical information is
extracted from this function the gauge dependence of many physical
quantities (in particular, the gauge dependence of the electron
spectra) remains completely unclear and each time the special
investigations are necessary to display it. Moreover all transformations
in [1-3] were made on the formal level ignoring the problems which
encounter the real calculations (e.g. the problem of treating
divergencies, in the first rate infrared divergencies). The real
calculations which were done later in [4,5] have demonstrated that
within this problem many questions have remained unsolved and require
the more careful investigations. For QFT in the recent paper [6] a
number of these questions were eliminated but the direct extension of
its results to statistical QED is doubtful and requires the special
revision: firstly the electron spectrum in statistical QED is
completely different from QFT (besides one-particle excitations there
are the collective ones with different effective masses); and secondly
in statistics unlike QFT  all possible degrees of freedom are
revived, therefore the special requirements are necessary to eliminate
their influence on the physical results when any regularization is
made.  Moreover all infrared divergencies are aggravated in statistics
and, for example, any calculations in the arbitrary $\alpha$-gauge
require many efforts to block up the additional divergencies from the
new pole in ${\cal D}^0$-function (the $1/p^4$-pole). Only the Feynman
and Coulomb gauges remain enough reliable since these gauges can be
exploited without any additional infrared regularization and there is a
guarantee that all exact theory properties will be kept: in the first
rate the gauge invariance which is very sensitive to any calculational
disagreements.

In this paper both gauges (the Feynman and Coulomb ones) are used to
determine the electron spectra in statistical QED and investigate
their gauge dependence. Sometimes the arbitrary $\alpha$-gauge will be
briefly discussed but only to illustrate the results found. Two
different kinds of the electron spectra in statistical QED are found:
the one-particle and collective excitations. The one-particle electron
spectrum which results from perturbative calculations is similar to the
bare electron one and to the spectrum found in QFT. The collective
electron excitations are generated by nonperturbative calculations and
in the leading order have not any analogy in QFT. This spectrum has
four well-separated branches which present the quasi-particle and
quasi-hole excitations [7,8,9]; all branches are always massive (there
is a well-separated spectrum of four effective masses) and their
properties are completely different from the bare one. To perform
calculations the standard temperature Green function technique is used
and the case of zero damping is only considered. We concentrate our
attention to solve the fermion dispersion relation in a more complete
form and to investigate the gauge dependence of all spectra found.

\section{ QED Lagrangian and electron self-energy}
The QED Lagrangian in covariant gauges has the form
\setcounter{equation}{0}
\begin{eqnarray}
{\cal L}=&-&\frac{1}{4}{F_{\mu\nu}}^2+{\bar \psi}[\gamma_{\mu}
(\partial_{\mu}-ie V_{\mu})+m]\psi \nonumber\\ &-&\mu{\bar
\psi}\gamma_4\psi +\frac{1}{2\alpha}(\partial_{\mu}V_{\mu})^2
\end{eqnarray}
where $F_{\mu\nu}=\partial_{\mu}V_{\nu}-\partial_{\nu}V_{\mu}$  is
the Abelian field strength;  $\psi$  and ${\bar \psi}$ are the
Dirac fields;  $\mu$ and $m$ are the electron chemical potential and
the bare electron mass, respectively;  $\alpha$ is the gauge fixing
parameter ($\alpha=1$ for the Feynman gauge). The metrics is chosen to
be Euclidean, and $\gamma_{\mu}^2=1$.

To find the Fermi excitations in hot QED we start with the usual
Schwinger-Dyson equation
\begin{eqnarray}
G^{-1}(q)=G_0^{-1}(q)+\Sigma(q)
\end{eqnarray}
and calculate the electron self-energy according to its exact
representation [3]
\begin{eqnarray}
\Sigma(q)=\frac{e^2}{\beta}\sum_{p_4}^B
\int\frac{d^3p}{(2\pi)^3}{\cal D}_{\mu\nu}(p)\gamma_{\mu}G(p+q)
\Gamma_{\nu}(p+q,q|p)\;.
\end{eqnarray}
We study $\Sigma(q)$ only in the one-loop approximation where the
bare Green functions and the bare vertex are used to calculate Eq.(3).
Using this expression for $\Sigma(q)$ the electron excitations will be
found in two cases:  perturbatively, exploiting the bare mass shell
(one-particle spectrum) and nonperturbatively when the spectrum and a
new mass shell are determined simultaneously (collective excitations).
Both these spectra (especially their long wavelength limits) will be
considered in different gauges (in the Feynman and Coulomb ones) to
understand their influence on the final result.

Within the one-loop approximation the exact decomposition for
$\Sigma(q)$ is given by
\begin {eqnarray}
\Sigma(q)=i\gamma_{\mu}K_{\mu}(q)+m\;Z(q)
\end{eqnarray}
and is used to find nonperturbatively the function G(q)
\begin{eqnarray}
G(q)=\frac{-i\gamma_{\mu}({\hat q_{\mu}}+K_{\mu})+m\;(1+Z)}
{({\hat q_{\mu}}+K_\mu)^2\;+\;m^2\;(1+Z)^2} \,.
\end{eqnarray}
This representation leads to the one-loop dispersion relation for the
Fermi excitations which in any gauge has the form
\begin{eqnarray}
[\;(iq_4-\mu)-{\bar K}_4]^2\;=\;\q^2\;(1+K)^2 +m^2(1+Z)^2
\end{eqnarray}
and after the standard analytical continuation it can be solved
analytically or numerically. Here $K_4=i{\bar K_4}$ and
${\hat q}=\{(q_4+i\mu),\q\}$.

The one-loop expression for $\Sigma(q)$ in the Feynman gauge was
used many times earlier and is known as follows [9]
\begin{eqnarray}
&&\!\!\!\!\!\!\!\!\!\!\!\!\!\!\!\Sigma^F(q)=
-e^2\int\frac{d^3p}{(2\pi)^3}\;\left\{\;\Bigr[
\frac{1}{\epsilon_\p}\;\frac{n_\p^+\;[\gamma_4\epsilon_\p+(i\g\p+2m)]}
{[q_4+i(\mu+\epsilon_\p)\;]^2+(\p-\q)^2}\right.\nonumber\\
&&\!\!\!\!\!\!\!\!\!\!\!\!\!\!\!\!\!\! \left.+\;\frac{n_\p^B}{|\p|}
\;\frac{(|\p|+\mu-iq_4) \gamma_4-[i\g(\p+\q)+2m]}{[q_4+
i(\mu+|\p|)\;]^2+\epsilon_{\p+\q}^2}\;\Bigr]
-\Big[h.c.;(m,\mu)\rightarrow-(m,\mu)\Big]\right\}
\end{eqnarray}
where $\epsilon_\p=\sqrt{\p^2+m^2}$ is the bare electron energy;
$n_\p^{B}=\left\{\exp\beta|\p|-1\right\}^{-1}$ and $n_\p^{\pm}=
\left\{\exp\beta(\;\epsilon_\p \pm \mu)+1\right\}^{-1}$ are the Bose
and Fermi occupation numbers, respectively.

The analogues expression for $\Sigma(q)$ in the Coulomb gauge is more
complicated and can be calculated within Eq.(3) using for this case the
appropriate ${\cal D}^0$-function
\begin{eqnarray}
{\cal D}^0_{ij}(p)\;=\;\Bigr(\delta_{ij}-\frac{p_i p_j}{\p^2}\Bigr)
\frac{1}{p^2_4+\p^2}  \,,\qquad {\cal D}^0_{44}(p)\;=\;\frac{1}{\p^2}
\end{eqnarray}
The standard $\gamma$-matrix algebra is used to make the summation over
spinor indices
\begin{eqnarray}
\Sigma^C&=&\frac{2e^2}{\beta}\sum_{p_4}^B
\int\frac{d^3p}{(2\pi)^3}\frac{m+i\gamma_4(p+{\hat q})_4+i\g\p
\;[(\p+\q|\p)/\p^2]}{[(p+{\hat q})^2+m^2]\; p^2}\nonumber\\
&+&\frac{e^2}{\beta}\sum_{p_4}^B\int\frac{d^3p}{(2\pi)^3}
\frac{m-i\gamma_4(p+{\hat q})_4+i\g(\p+\q)}
{[(p+{\hat q})^2+m^2]\; \p^2}
\end{eqnarray}
and then the remaining summation over the Bose frequencies within
Eq.(9) (or over the Fermi ones in its another form) should be made in
the usual manner [3].  The result has the form
\begin{eqnarray}
\Sigma^C(q)&=&-e^2\int\frac{d^3p}{(2\pi)^3}\;\left\{\;\Bigr[
\frac{n_\p^+}{\epsilon_\p}\;
\frac{\gamma_4\epsilon_\p+i\g(\p-\q)\;[(\p-\q|\p)/(\p-\q)^2]+m}
{[q_4+ i(\mu+\epsilon_\p)\;]^2+(\p-\q)^2}\;\right.\nonumber\\
&+&\left.\;\frac{n_\p^B}{|\p|}\;
\frac{(|\p|+\mu-iq_4)\gamma_4-i\g\p\;[(\p|\p+\q)/\p^2]-m}
{[q_4+ i(\mu+|\p|)\;]^2+\epsilon_{\p+\q}^2}\;\right.\nonumber\\
&-&\left.\;\frac{n_\p^+}{\epsilon_\p}\;
\frac{\gamma_4\epsilon_\p-i\g\p-m}{2(\p-\q)^2}\;\Bigr]
-\Big[h.c.;(m,\mu)\rightarrow-(m,\mu)\Big]\right\}
\end{eqnarray}
and along with Eq.(7) will be used to calculate the electron spectrum
and its limits.

\section{One-particle electron spectrum}
Now we investigate the one-particle electron excitations in the
different gauges: at first in the Feynman and Coulomb gauges and then
in the arbitrary $\alpha$-gauge. The spectrum found (in any case its
long wavelength limit) is gauge invariant and qualitatively is the same
as the bare one. It has two branches $iq_4=\mu_R \pm m_R$ but their
chemical potential and effective mass are quantitatively improved due
to interaction with the medium.  To find this spectrum in the leading
order the dispersion relation (6) is solved with $\Sigma(q_4,\q)$ taken
at once on the bare mass shell $iq_4=\mu \pm \sqrt{\q^2+m^2}$ that is
evidently correct to establish $e^2$-corrections. However this procedure
requires the additional analysis to establish the next-to-leading term
(e.g. the $e^4$-term) where the perturbative calculations are more
complicated. In this case the mass shell should be shifted as well to
make all calculations selfconsistently.

The one-particle electron spectrum in the leading order is:
\begin{eqnarray}
iq_4=(\mu+{\bar K}_4) \pm
\sqrt{m^2\Bigr(1+Z(\q)\Bigr)^2+\q^2\Bigr(1+K(\q)\Bigr)^2}
\end{eqnarray}
since all functions being put at once on the bare mass shell are
independent from $iq_4$. Our task is to calculate (11) explicitly using
the expressions for $\Sigma(q)$ found above in the Feynman (F.G) and
Coulomb (C.G) gauges. These calculations are rather lengthy since they
require at first to extract the functions $Z(q)$ and $K_\mu(q)$ from
$\Sigma(q)$ (e.g. as it is done in the Coulomb gauge)
\begin{eqnarray}
&&Z^C(q)=-\epsilon^2 \int\frac{d^3p}{(2\pi)^3}\;
\left\{\;\Bigr[\;\frac{n_\p^+}{\epsilon_\p}\;\Bigr(
\frac{1}{[q_4+i(\mu+\epsilon_\p)\;]^2+(\p-\q)^2}
+\frac{1}{2(\p-\q)^2}\Bigr) \right.\nonumber\\
&&-\left.\frac{n_\p^B}{|\p|}\;\;\frac{1}
{[q_4+i(\mu+|\p|)\;]^2+\epsilon_{\p+\q}^2}\;
\Bigr]\;+\;\Big[\;h.c.;(\mu\rightarrow-\mu)\;\Bigr]\;\right\}
\nonumber
\end{eqnarray}
\begin {eqnarray}
&&iK^C_4(q)=-\epsilon^2\int\frac{d^3p}{(2\pi)^3}\;
\left\{\;\Bigr[\;n_\p^+\;\Bigr(\frac{1}{[q_4+i(\mu
+\epsilon_\p)\;]^2+(\p-\q)^2}
-\frac{1}{2(\p-\q)^2}\Bigr)\right.\nonumber\\
&&\left.+\frac{n_\p^B}{|\p|}\;\frac{|\p|+\mu-iq_4}
{[q_4+i(\mu+|\p|)\;]^2+\epsilon_{\p+\q}^2}\;\Bigr]\;
-\;\Bigr[\;h.c.;(\mu\rightarrow-\mu)\;\Bigr]\;\right\}
\end{eqnarray}
and then, after the bare mass shell has been put into these expressions,
to expand them in the powers of $\p$. However, in the leading order
which is only considered here, the final result is very simple and can
be presented as follows
\begin{eqnarray}
&&{\bar K}_4(0)\;=\;-\;2\;I_B \pm \frac{I_A}{m}\; \,,
\qquad Z(0)\;=\;-\;4\;I_Z \pm 4\;\frac{I_B}{m} \qquad (F.G)\nonumber\\
&&{\bar K}_4(0)\;=\;\pm \frac{I_A}{m}\; \,,\qquad\;\;\;\;\;
\qquad Z(0)\;=\;-\;4\;I_Z \pm2\;\frac{I_B}{m}\; \qquad  (C.G)
\end{eqnarray}
to demonstrate that these functions are gauge dependent (the same as
$\Sigma(q)$ taken on the bare mass shell [4,5]).

Nevertheless the long wavelength limit for the one-particle electron
excitations (due to the algebraic transformations) has another form
\begin{eqnarray}
iq_4=\mu_R\;\pm\;m_R=\mu\;(\;1\;+\;2\; {\tilde I}_B\;)
\;\pm\;\Bigr[\;m(1-4I_Z)+\frac{I_A}{m}\;\Bigr]
\end{eqnarray}
and this limit is the same in both gauges. Here $I_B=\mu{\tilde I}_B$
and other abbreviations are:
\begin {eqnarray}
&&I_A\;=\;e^2\int\limits_0^{\infty}\frac{d|\p|}{2\pi^2}
\;\Bigr[\;\epsilon_\p
\,\frac{n_\p^++n_\p^-}{2}\;+\;|\p|\,n_\p^B\;\Bigr] \,,\nonumber \\
&&I_B=-e^2\int\limits_0^{\infty}
\frac{d|\p|}{4\pi^2}\;\frac{n_\p^+-n_\p^-}{2} \,,\qquad
I_Z=e^2\int\limits_0^{\infty}
\frac{d|\p|}{4\pi^2}\;\frac{n_\p^++n_\p^-}{2\epsilon_\p}\;.
\end{eqnarray}
The found invariance is very specific and, undoubtedly, is destroyed when
the spectrum for all momenta is considered. However, the effective
electron mass (the one-particle spectrum limit at zero momentum
found perturbatively) is gauge invariant, and namely this fact is
demonstrated above.

The obtained spectrum limit can be repeated in the arbitrary
$\alpha$-gauge to demonstrate the difficulties which arise due to the
additional pole $1/p^4$ and their influence on the calculations made.
In this gauge the ${\cal D}^0$-function is more complicated and has the
 additional tensor structure
\begin{eqnarray}
{\cal D}^0_{ij}(p)\;=\;\frac{\delta_{\mu\nu}}{p^2}\;+\;(\alpha-1)\;
\frac{p_\mu p_\nu}{p^4}
\end{eqnarray}
which introduces at once the $\alpha$-dependence into all calculations.
Now the one-loop electron self-energy (3) is transformed to be
\begin{eqnarray}
\Sigma(q)\;=\;\Sigma(q)^F\;+\;(\alpha-1)\;\frac{e^2}{\beta}\sum_{p_4}^B
\int\frac{d^3p}{(2\pi)^3}(\gamma_{\mu}p_\mu)\frac{G(p+q)}{p^4}
(\gamma_{\nu}p_\nu)
\end{eqnarray}
where the first term reproduces the standard result for $\Sigma(q)$ in
the Feynman gauge  and the last one gives its gauge dependence.
This term can be calculated in the more explicit form using (for
example) the Ward identity for statistical QED [3]
\begin{eqnarray}
i q_\mu \Gamma_\mu(p,p+q|q)\;=\;G^{-1}(p+q)\;-\;G^{-1}(p)
\end{eqnarray}
and the simple algebra. The final result has the known form
\begin{eqnarray}
\Sigma(q)^\alpha\;=\;(\alpha-1)\;\frac{e^2}{\beta}\sum_{p_4}^B
\int\frac{d^3p}{(2\pi)^3}(\gamma_{\mu}p_\mu)\frac{G(p+q)}{p^4}
(\gamma_{\nu}p_\nu)\;=\;G^{-1}(q)\Sigma'(q)
\end{eqnarray}
which is very convenient for our discussion if there is a possibility
to regularize the $\Sigma'(q)$-quantity.  Without any additional
regularization this quantity is singular on the bare mass shell and
$\alpha$-dependence is kept (as it was found above here and in [4,5]).
However, if any regularization is done to make $\Sigma'(q)$ finite, the
electron self-energy taken on the bare mass shell is gauge invariant
and reproduces the spectrum found in the Feynman gauge.  Moreover this
spectrum will be gauge invariant for any momenta and, namely, this fact
in comparison with the results obtained above, is very doubtful.  The
most probably that such regularization of $\Sigma'(q)$ is possible only
for $\q=0$ to be in agreement with the results of calculations in the
Feynman  and Coulomb gauges. However, the more complicated situation is
not excluded: the independence for $\alpha$ does not mean else the real
gauge invariance. To be sure one needs to check this fact in another
outstanding gauge which is free on any additional regularization (e.g.
the Coulomb gauge attracts the special attention). Probably, within
Eq.(19) namely this situation is taken place, but formally here any
regularization is acceptable since its influence on the one-particle
electron spectrum is blocked up due to the zero factor which is
$G^{-1}(p)$ on the bare mass shell.

\section{Collective electron spectrum in Coulomb gauge}
This gauge along with the Feynman gauge (where $\alpha=1$) is very
convenient for the practice calculations since no additional infrared
regularization is necessary to treat all perturbative diagrams.
This is not the case for any $\alpha$-gauge where a new pole
($1/p^4$-pole in the ${\cal D}_0$-function) introduces many
additional difficulties and makes all calculations very complicated.
Indeed in the $\alpha$-gauge the one-loop electron self-energy is
given by
\begin{eqnarray}
\Sigma^\alpha(q)&=&\frac{e^2}{\beta}\sum_{p_4}^B
\int\frac{d^3p}{(2\pi)^3}\frac{(\alpha+3)m+(\alpha+1)i\gamma_\mu
(p+{\hat q})_\mu}{[p+{\hat q})^2+m^2]p^2}\nonumber\\
&-&(\alpha-1)\;\frac{e^2}{\beta}\sum_{p_4}^B
\int\frac{d^3p}{(2\pi)^3}\frac{2i\gamma_\mu p_\mu (p+{\hat q}|p)}
{[p+{\hat q})^2+m^2] p^4}
\end{eqnarray}
and one can see that now the collective spectra which will be found
exploiting this expression is surely gauge dependent. Moreover, since
$Z^{\alpha}= Z^F(\alpha+3)/4$, this dependence is kept for the long
wavelength spectrum limit as well. Moreover the calculation of Eq.(20)
requires the special procedure to treat its last integral: for example,
one can consider that
\begin{eqnarray}
\frac{1}{p^4}=\raisebox{-2.5mm}{$\buildrel{\displaystyle\lim}\over
{\scriptstyle \kappa^2\to 0}$}\Bigr(-\frac{\partial}{\partial\kappa^2}
\bigr)\frac{1}{p^2+\kappa^2}
\end{eqnarray}
or introduce another way of calculation. It is not excluded
that the additional infrared regularization ( e.g. the dimensional
regularization) will be necessary.  In any case the result will be
rather complicated and non-single value.

The situation is different in the Coulomb gauge where all calculations
are standard and selfconsistent. Their result was done above and will be
used here to solve the dispersions relation (6). Only the long
wavelength spectrum limit (the case $\q=0$ in Eq.(6)) will be
considered below and therefore only two functions $Z$ and $K_4$-should
be extracted from Eq.(10). These functions are given by Eq.(12) and can
be algebraically transformed to be
\begin {eqnarray}
&&Z^C(q_4,0)\;=\;-e^2\;\int\limits_0^{\infty}
\frac{d|\p|}{4\pi^2}\;\left\{\;\frac{4\p^2}
{4\epsilon_\p^2(iq_4-\mu)^2-[(iq_4-\mu)^2+m^2]^2}\right.\nonumber\\
&&\left.\Bigr[\;\frac{(iq_4-\mu)^2+m^2}
{\epsilon_\p}\;\frac{n_\p^++n_\p^-}{2}
\;+\;2(iq_4-\mu)\;\frac{n_\p^+-n_\p^-}{2}\Bigr]\right.\nonumber\\
&&\left.-\;n_\p^B\;\frac{4|\p|\;[(iq_4-\mu)^2-m^2]}
{4|\p|^2(iq_4-\mu)^2-[(iq_4-\mu)^2-m^2]^2}\;+
\;\frac{n_\p^++n_\p^-}{\epsilon_\p} \right\}
\end{eqnarray}
and then analogously we find $K_4$-function
\begin {eqnarray}
&&\!\!\!\!\!\!iK^C_4(q_4,0)\;=\;-e^2\;\int
\limits_0^{\infty}\frac{d|\p|}{2\pi^2}\left\{\frac{4\p^2\epsilon_\p}
{4\epsilon_\p^2-(iq_4-\mu)^2[1+{\displaystyle\frac{m^2}
{(iq_4-\mu)^2}}]^2}\right.\nonumber\\
&&\!\!\!\!\!\!\left.\Bigr[\;\frac{n_\p^++n_\p^-}{2}\;
+\;\frac{(iq_4-\mu)}{2\epsilon_\p}\Bigr(1+\frac{m^2}
{(iq_4-\mu)^2}\Bigr)\frac{n_\p^+-n_\p^-}{2}\Bigr]
-(iq_4-\mu)\frac{n_\p^+-n_\p^-}{2}\right.\nonumber\\
&&\!\!\!\!\!\!\left.+n_\p^B\;\frac{4\p^3}{4\p^2-(iq_4-\mu)^2
[1-{\displaystyle\frac{m^2}{(iq_4-\mu)^2}}]^2}\Bigr[\;1-\frac{(iq_4
-\mu)^2-m^2}{2|\p|^2}\Bigr]\right\}\;\frac{1}{iq_4-\mu}
\end{eqnarray}
Now our problem is to solve Eq.(6) explicitly. To this end we should
put Eq.(22)-(23) on the new mass shell $(iq_4-\mu)=\omega$ and in this
form substitute these integrals into Eq.(6). However in this case the
arisen equation is very complicated and can be used only for numerical
calculations. On the other hand, keeping the perturbative accuracy
the obtained integrals could be simplified using a condition $m<<T$
and omitting all $m^2$ terms. If only the leading term for the small $m$
is kept, these functions have the form
\begin{eqnarray}
-iK^C_4(q_4,0)\;=\;\frac{I_A}{\omega}\;+\;I_B\\
Z^C(q_4,0)\;=\;-3I_Z\;+\;2\frac{I_B}{\omega}
\end{eqnarray}
where all integrals were earlier determined by Eq.(15). The dispersion
equation is the simple quadratic equation
\begin{eqnarray}
\omega^2-\omega [\;\eta\; m (1-3I_Z)+I_B\;]-(\;I_A+2\eta mI_B\;)\;=\;0
\end{eqnarray}
whose solution reproduces the final result. This result is given by
\begin{eqnarray}
E^C(0)=\mu+\frac{1}{2}\Big[\eta\;m_R+I_B\Big]\pm
\sqrt{\;\frac{[\eta\;m_R+I_B]^2}{4} +(I_A+2\eta m I_B)} \,.
\end{eqnarray}
where $m_R=m(1-3I_Z)$ and at once Eq.(27) demonstrates that the
spectrum found in the Coulomb gauge is qualitatively the same as the
one in the Feynman gauge. In particular, the spectrum branches are
split even at zero momentum and have four well-separated effective
masses:  two of them are related to the quasi-particle excitations and
two others present the new quasi-hole ones. This is evident from
Eq.(27) where $\eta=\pm 1$.  However, this spectrum is not coincident
quantitatively with the one found in the Feynman gauge [9]
\begin{eqnarray}
E^F(0)=\mu+\frac{1}{2}\Big[\eta\;m_R-I_B\Big]\pm
\sqrt{\;\frac{[\eta\;m_R-I_B]^2}{4} +(I_A+4\eta m I_B)} \,.
\end{eqnarray}
where $m_R=m(1-2I_Z)$ and the $\mu$-dependence are different. Only for
the case $m,\mu$ these results are similar and gauge independent.

\section{Conclusion}
To summarize we have established two kinds of the electron excitations
in statistical QED: the one-particle spectrum and collective one which
in the leading order is absent in QFT. The one-particle spectrum is
qualitatively the same as the bare one and can be gauge invariant, in
any case its effective mass is always the gauge invariant quantity. It
has two branches $iq_4=\mu_R \pm m_R$ and their chemical potential and
effective mass are quantitatively improved due to the interaction with
the medium. This is not the case for the collective electron spectrum
which has nonperturbative nature and is completely different from the
bare one. This spectrum is split and their branches develop the
different dynamical masses (the gap at zero momentum) which are always
nonzero in the medium. These dynamical masses are not connected with
bare masses and are generated always even for the case $m,\mu=0$ being
in the last case the gauge invariant quantity. But if the bare electron
mass is not equal to zero all parameters which determine the
collective spectrum are gauge dependent and their connection with the
real electron excitations should be investigated separately. However
one can see (comparing our results in the Feynman and Coulomb gauges)
that these changes are only quantitative and qualitatively the found
spectrum is the same in any gauges and namely that, probably,
demonstrates the physical sense of the gauge invariance principle. The
same situation is not new and takes place, for example, with the
infrared fictitious pole in statistical QCD [10] which can be found in
any gauges but quantitatively its position depends on the gauges fixed.
Many of the results found here are valid as well for statistical QCD in
the leading order but in the non-Abelian theory  the situation is more
complicated and surely requires the understanding of the gauge
invariance principle in a broader sense.

\newpage

\begin{center}
{\bf References}
\end{center}

\renewcommand{\labelenumi}{\arabic{enumi}.)}
\begin{enumerate}

\item{ L.~D.~Landay and I.~M.~Khalatnikov, Sov. Phys. JETP {\bf 2}
(1955) 69.}

\item{ E.~S.~Fradkin, Sov. Phys. JETP {\bf 2} (1955) 258.}

\item{ E.~S.~Fradkin, The doctoral thesis from FIAN (1960); published
in Trudy Fiz. Inst. {\bf 29} (1965) 7 (Proc. P.~N.~Lebedev Physical
Inst. {\bf 29} (1967) 1).}

\item{ R.~Jackiw and S.~Templeton, Phys. Rev. {\bf D23} (1981) 2291.}

\item{ S.~Deser, R.~Jackiw and S.~Templeton, Ann.Phys. (N.Y) {\bf 140}
(1982) 372.}

\item{ I.~V.~Tyutin and Vad.~Yu.~Zeitlin, hep-th/9711137.}

\item{ V.~V.~Klimov, Yad.Fiz. {\bf 33} (1981) 1734 (Sov. J. Nucl. Phys.
{\bf 33} (1981) 934); Zh. Eksp. Teor. Fiz. {\bf 82} (1982) 336 (Sov.
Phys. JETP  {\bf 55} (1982) 199) .}

\item{ R.~D.~Pisarski, Nucl. Phys. {\bf A498} (1989) 423c.}

\item{ O.~K.~Kalashnikov, Mod. Phys. Lett. {\bf A12} (1997) 347 and
also  Pis'ma  Zh. Eksp. Teor. Fiz. {\bf 67} (1998) 3 (JETP Lett.
{\bf 67} (1998) 1).}

\item{ O.~K.~Kalashnikov, Phys. Lett. {\bf B 279} (1992) 367. }

\end{enumerate}

\end{document}